\documentclass[10pt,conference,letterpaper]{IEEEtran}

\usepackage[T1]{fontenc}
\usepackage{cite}
\usepackage{amsmath,amssymb}
\usepackage{graphicx}
\usepackage{booktabs,array}
\usepackage{multirow}
\usepackage{placeins}
\usepackage{balance}
\usepackage{xcolor}
\usepackage[hidelinks]{hyperref}

\newif\ifdraftnotes
\draftnotesfalse

\newcommand{\yes}{\ensuremath{\checkmark}}
\newcommand{\partialsupport}{\ensuremath{\triangle}}

\title{TriMedium: A Cross-Medium Simulator for Embodied Networking Across Air-Ground-Sea}

\author{%
\IEEEauthorblockN{Ruizhe Dang\textsuperscript{1}, Jiayi Su\textsuperscript{1},
Wenlihan Lu\textsuperscript{1}, Weijie Yuan\textsuperscript{2}, and Shijian Gao\textsuperscript{1}}
\IEEEauthorblockA{%
\textsuperscript{1}\textit{The Hong Kong University of Science and Technology (Guangzhou)}, Guangzhou, China\\
\textsuperscript{2}\textit{Southern University of Science and Technology}, Shenzhen, China}
}

\begin{document}
\bstctlcite{IEEEexample:BSTcontrol}
\maketitle

\begin{abstract}
Embodied intelligent applications have expanded from air and ground into
maritime environments, where missions such as coastal reporting require
coordination among underwater sensing, surface gateways, and aerial
relays. Yet field trials remain costly, and existing simulation tools make
it difficult to study how cross-medium communication affects cooperation
among these platforms. We present TriMedium, an integrated Air-Ground-Sea simulator
for cross-medium embodied networking. TriMedium has three design elements.
First, it provides a shared three-dimensional world where aerial,
terrestrial, and underwater entities coexist, so underwater agent motion
and visual observations stay geometrically consistent with air and ground
nodes. Second, it uses a common simulation clock that preserves observation
timestamps while scheduling delayed communication and computation,
clarifying when the shore receiver obtains each observation. Third,
modular quantization and compression interfaces accommodate heterogeneous
telemetry and image pipelines, while integrity checks separate intentional
lossy distortion from incomplete or corrupted delivery. Three case studies
validate TriMedium. Multimodal data delivery tests telemetry and image
pipelines, scene construction examines cross-domain spatial consistency,
and positioning assesses geometric accuracy under delayed communication.
Together, they show TriMedium enables embodied networking by quantifying
cross-medium delivery dynamics and fidelity across heterogeneous payloads.

\end{abstract}


\section{Introduction}
\label{sec:introduction}

Embodied intelligent applications have surged in both air and ground
domains, driven by two major application fronts: the low-altitude economy,
where UAVs are expanding into logistics, inspection, and environmental
monitoring~\cite{zeng2016uav}, and autonomous driving, which is transforming
terrestrial mobility. Such applications rely on embodied intelligence, in
which agents perceive their surroundings and act within physical
environments, and their cooperation requires exchanging observations and
state information over wireless networks~\cite{liang2026synergetic}.
Air--ground cooperative systems connect these capabilities, and platforms
such as CARLA-Air~\cite{carlaair2026} already provide a shared simulation
environment for studying this cooperation. Growing operational demand is
now pushing these applications into maritime environments.

\begin{figure}[!htbp]
\centering
\includegraphics[width=\columnwidth]{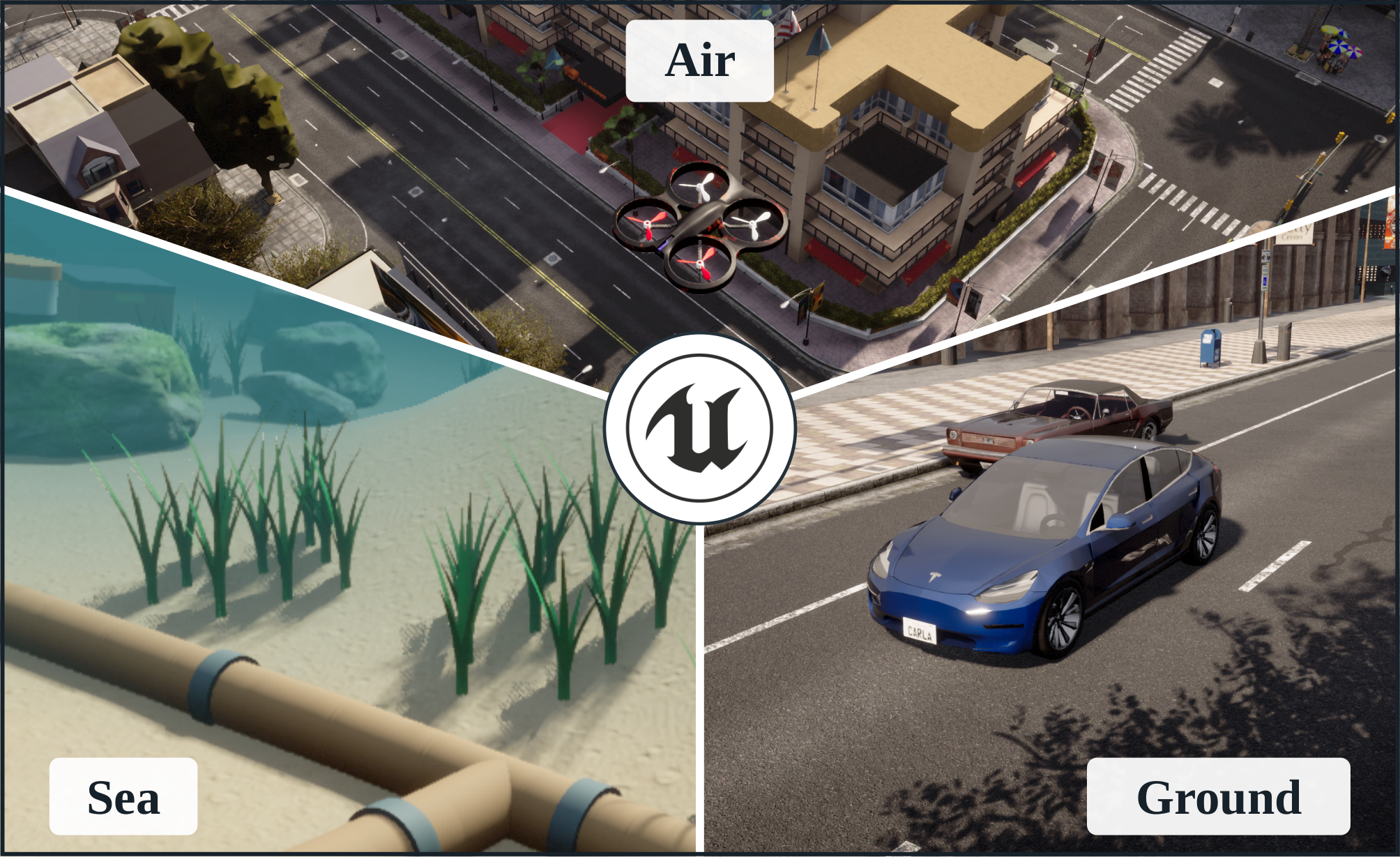}
\caption{Air--Ground--Sea environments in TriMedium, illustrated by
aerial, terrestrial, and underwater scene views.}
\label{fig:three-domain-overview}
\end{figure}

Coastal monitoring illustrates this trend: observations acquired by an
autonomous underwater vehicle (AUV) must reach shore facilities through a
surface gateway and an aerial relay, so that information traverses acoustic
and radio-frequency (RF) links across air, ground, and sea. Such missions
couple platform motion and sensing with cross-medium data
delivery~\cite{gao2026lowaltitudeiscc}. Investigating them in the field
requires vehicle deployment and recovery, coordination among platforms, and
logistical support, which make repeated testing costly and difficult to
reproduce~\cite{zhang2022dave}; simulation therefore provides a practical
basis for examining different configurations within the same cooperative
scenario.

Marine robotics simulators such as HoloOcean~\cite{potokar2022holoocean}
and Stonefish~\cite{cieslak2019stonefish} support underwater motion,
sensing, and acoustic communication. SimART~\cite{yan2026simart} connects
robotic observations with RF modeling and evaluation.
Sionna RT~\cite{hoydis2023sionnart} supports scene-based radio propagation
modeling, while ns-3~\cite{riley2010ns3} provides acoustic and RF network
models. Yet these platforms offer limited joint support for
Air--Ground--Sea embodied activity and acoustic--RF delivery across
observation, processing, cross-medium forwarding, and reconstruction.
Table~\ref{tab:platforms} compares representative platforms against these
requirements. Addressing this gap raises three design challenges:
maintaining spatial and temporal consistency among platform states,
observations, and communication events; supporting replaceable processing
methods while consistently accounting for their payload sizes, processing
delays, and reconstruction results; and distinguishing intentional lossy
distortion from incomplete or corrupted delivery.

\begin{table*}[t]
\caption{Comparison of representative simulation platforms for
air--ground--sea embodied networking.}
\label{tab:platforms}
\centering
\footnotesize
\begin{tabular*}{\textwidth}{@{\extracolsep{\fill}}l*{9}{c}@{}}
\toprule
Platform & \multicolumn{3}{c}{Domain coverage}
& \multicolumn{2}{c}{Embodied simulation}
& \multicolumn{4}{c}{Networking and processing} \\
\cmidrule(lr){2-4}\cmidrule(lr){5-6}\cmidrule(l){7-10}
& \shortstack{Aerial}
& \shortstack{Terrestrial}
& \shortstack{Under-\\water}
& \shortstack{Visual\\sensing}
& \shortstack{Joint\\simulation}
& \shortstack{Acoustic\\links}
& \shortstack{RF\\links}
& \shortstack{Cross-medium\\relaying}
& \shortstack{Comp. /\\quant.} \\
\midrule
CARLA~\cite{dosovitskiy2017carla}
  & & \yes & & \yes & & & \yes & & \\
AirSim~\cite{shah2018airsim}
  & \yes & \yes & & \yes & & & & & \\
HoloOcean~\cite{potokar2022holoocean}
  & \yes & \yes & \yes & \yes & \yes & \yes & & & \\
Stonefish~\cite{cieslak2019stonefish}
  & \partialsupport & \partialsupport & \yes & \yes & \partialsupport & \yes & & & \\
UNav-Sim~\cite{amer2023unav}
  & & & \yes & \yes & & & & & \\
SimART~\cite{yan2026simart}
  & \yes & \yes & & \yes & & & \yes & & \\
ns-3~\cite{riley2010ns3}
  & \partialsupport & \partialsupport & \partialsupport & & & \yes & \yes & & \\
CARLA-Air~\cite{carlaair2026}
  & \yes & \yes & & \yes & \yes & & \yes & & \\
\textbf{TriMedium (ours)}
  & \yes & \yes & \yes & \yes & \yes & \yes & \yes & \yes & \yes \\
\bottomrule
\end{tabular*}
\par\smallskip
\begin{minipage}{\textwidth}
\footnotesize
\textit{Legend:} \yes\ = documented support;
\partialsupport\ = explicitly limited support;
blank = not established in the examined sources.
\end{minipage}
\end{table*}

To address these challenges, we present TriMedium\footnote{Project page:
\url{https://github.com/RuizheDang/TriMedium}.}, an integrated
Air--Ground--Sea simulator for cross-medium embodied networking.
TriMedium couples three layers. A shared
three-dimensional world places aerial, terrestrial, and underwater entities
in a common coordinate frame, so AUV trajectories, camera observations, and
air and ground node states remain geometrically consistent. A common
simulation clock records each observation's acquisition time and schedules
encoding, queueing, multi-hop transmission, and reconstruction according to
the configured processing and link models, clarifying which observation
reaches the receiver and when. Replaceable quantization, compression, and
reconstruction interfaces report encoded payload size and processing delay
back to the simulator, while acoustic and RF link models carry payloads
across medium boundaries and integrity checks separate transport errors
from intentional lossy distortion. Experiments are thus configured as a
single pipeline, with per-stage records covering observation time, encoded
volume, delay, delivery completeness, and reconstruction quality.

The main contributions made through the simulator development are as follows:

\begin{itemize}
  \item \textbf{Shared World and Synchronized Simulation:}
  A shared three-dimensional world places aerial, terrestrial, and
  underwater entities in a common spatial reference, while a common
  simulation clock preserves observation timestamps and schedules delayed
  communication and computation.

  \item \textbf{Extensible and Verifiable Processing
  Interfaces:} Replaceable quantization, compression, and reconstruction
  interfaces allow alternative methods to reuse the same scene,
  communication models, and evaluation workflow, while integrity checks
  separate intentional lossy distortion from incomplete or corrupted
  delivery.

  \item \textbf{Multi-Scenario Validation for Embodied Networking:} Three case
  studies on multimodal data delivery, scene construction, and positioning
  validate the simulator's use for embodied networking and quantify
  cross-medium delivery dynamics and fidelity across heterogeneous
  payloads.
\end{itemize}

\section{Platform Architecture}
\label{sec:platform}

\begin{figure*}[!t]
\centering
\includegraphics[width=\textwidth]{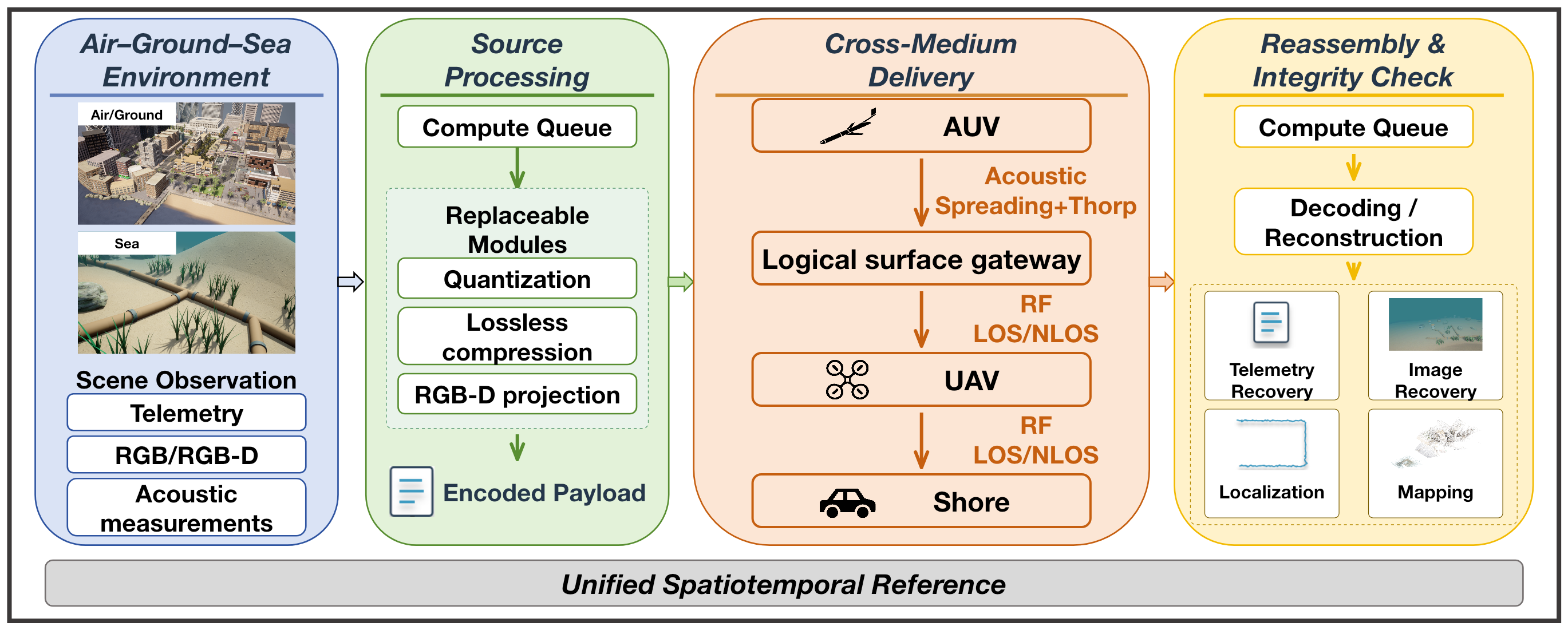}
\caption{Overview of TriMedium. Observations from the Air--Ground--Sea
environment pass through replaceable processing modules and cross-medium
relay paths to receiver-side verification and reconstruction. A unified
spatiotemporal reference coordinates agent activity, data processing, and
delivery, supporting telemetry reporting, image recovery, AUV positioning,
and Air--Ground--Sea mapping. Link models are described in
Section~\ref{sec:link-modeling}.}
\label{fig:platform-overview}
\end{figure*}

TriMedium simulates agent motion and sensing in a shared
Air--Ground--Sea environment, generating observations tied to agent poses
and the surrounding scene (Fig.~\ref{fig:platform-overview}). Replaceable
quantization and compression precede acoustic--RF delivery and
receiver-side reconstruction, with computational costs accounted for as
processing delays. A unified spatiotemporal reference, comprising a common
world coordinate frame and simulation clock, coordinates agent activity
with this data flow and determines when observations become available to
receiving tasks.

\subsection{Air--Ground--Sea Embodied Environment}
\begin{figure}[!t]
\centering
\includegraphics[width=\columnwidth]{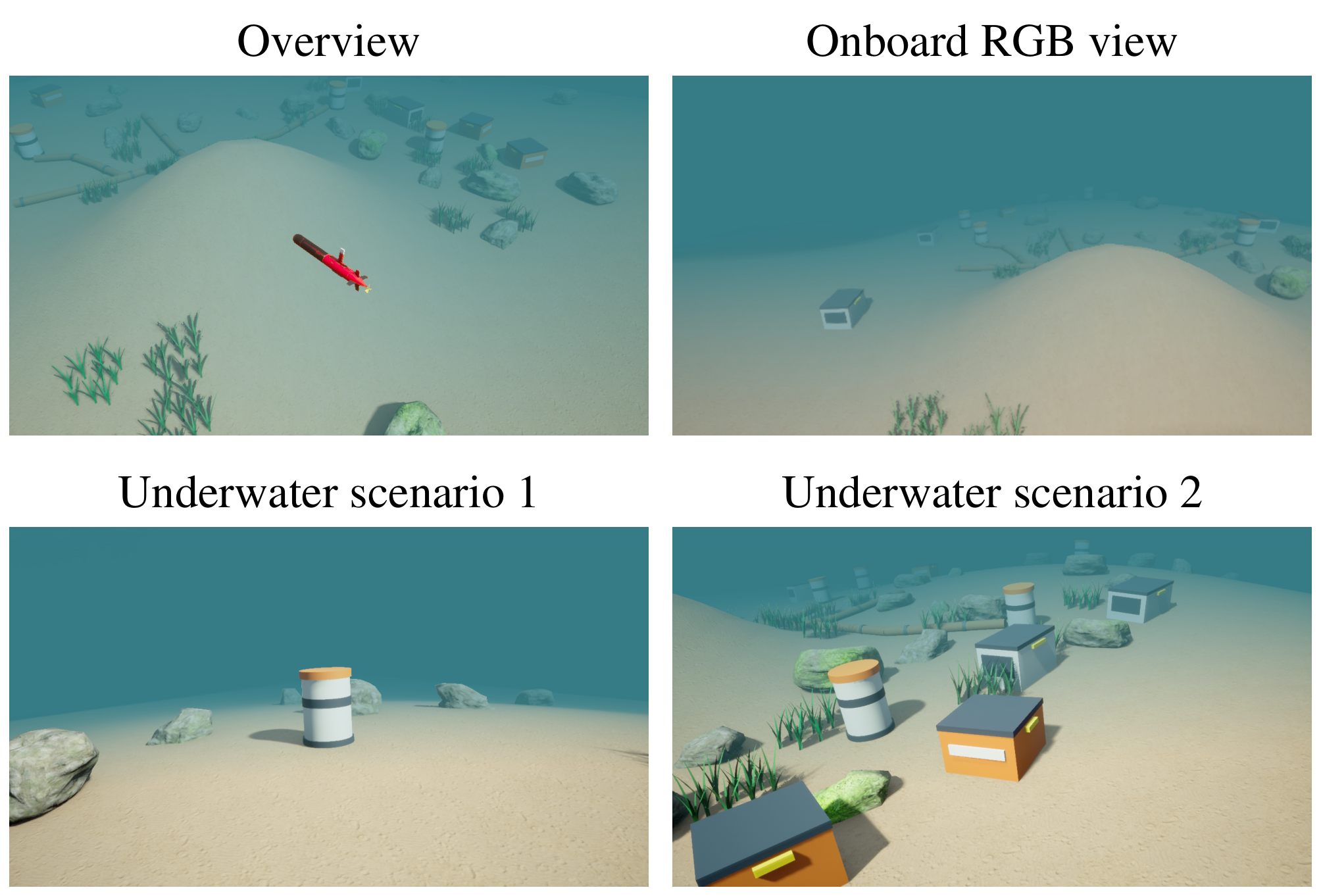}
\caption{Underwater environment and sensing in TriMedium.
Top row: scene overview with an AUV and its onboard RGB observation.
Bottom row: landmark and multiple-landmark views.}
\label{fig:underwater-environment}
\end{figure}

TriMedium brings aerial, ground, and underwater entities into a shared
three-dimensional environment. The implementation reuses the aerial and
ground simulation infrastructure of CARLA-Air~\cite{carlaair2026} and
incorporates underwater environments and AUV components through SeaPlugin.
The underwater environment contains undulating seabed terrain and
submerged objects. Mobile AUVs acquire observations as they traverse the
scene.
Fig.~\ref{fig:underwater-environment} illustrates this environment and
representative camera views. Entity positions and orientations across all
three domains are expressed in a common world coordinate frame.

TriMedium provides pose, velocity, and other vehicle states, together with
visual observations across the air, ground, and sea domains. In the
underwater environment, AUV cameras acquire RGB images, while the rendering
depth buffer provides corresponding scene depth for point-cloud generation
and mapping. State records and sensor outputs retain their acquisition
timestamps, and images are associated with the corresponding camera poses.
These data enter the subsequent processing and communication workflow for
use by receiving tasks.

\subsection{Extensible and Verifiable Processing Interfaces}
TriMedium supports state telemetry and image observations with optional
quantization to reduce precision and lossless compression to reduce
redundancy. Direct transmission retains the original representation.

Uniform quantization and zlib are replaceable reference implementations.
Researchers can introduce alternative methods and their corresponding
receiver-side recovery operations. Actual encoded payload sizes determine transmission demand, while
configured processing times contribute to delivery latency. These support
comparisons of transmitted volume, delay, and reconstruction quality.

Payloads exceeding the message-size limit are split into chunks. The
receiver reassembles them using chunk indices and counts, verifies total
length and hash against sender-provided values, and then decompresses and
reconstructs the data. Delivery completeness and reconstruction quality
are recorded separately, distinguishing incomplete or corrupted delivery
from intentional quantization distortion.

\begin{table*}[!t]
\caption{Experimental configuration of the cross-medium data delivery case study.}
\label{tab:settings}
\centering
\footnotesize
\setlength{\tabcolsep}{6pt}
\renewcommand{\arraystretch}{1.12}
\begin{tabular*}{\textwidth}{@{\extracolsep{\fill}}lll@{}}
\toprule
\textbf{Parameter} & \textbf{State telemetry} & \textbf{Image observations} \\
\midrule
Source data & Position, velocity, depth (7 FP32 values)
  & $640\times360$ PNG images \\
Sampling frequency & 20~Hz & 5~Hz \\
Reporting policy & 32-sample groups & Frame selection every 12~s \\
Workload size & 569 samples (18 groups) & 10 images \\
Direct transmission & FP32 & PNG \\
Lossless compression & zlib & zlib \\
Quantization + compression & 16-/8-bit uniform quantization + zlib
  & 6-/4-bit uniform quantization + zlib \\
Maximum acoustic rate & 32~kbit/s & 32~kbit/s \\
Maximum RF rate & 1~Mbit/s per link & 1~Mbit/s per link \\
\bottomrule
\end{tabular*}
\par\smallskip
\begin{minipage}{\textwidth}
\footnotesize
The final telemetry group contains 25 samples. For image observations,
zlib compression is applied to decoded pixel data.
\end{minipage}
\end{table*}

\subsection{Cross-Medium Embodied Networking}
TriMedium connects underwater, surface, aerial, and ground nodes through
acoustic and RF links. Surface gateways forward acoustically received
payloads into the RF network, where relays carry them toward their
destinations~\cite{alqurashi2023maritime}. These paths carry encoded data
from the processing interfaces to receiver-side reconstruction, allowing
observations acquired by one platform to support tasks at another.

During delivery, agents continue moving and acquiring new observations.
TriMedium coordinates processing, queueing, forwarding, and reconstruction
with scene activity through a common simulation
clock~\cite{wubben2023cosimulation}. Each observation retains its
acquisition time, and the receiver records when reconstruction finishes.
Receiving tasks can therefore determine both when information becomes
usable and which earlier scene state it describes.

\subsection{Link Modeling Rationale}
\label{sec:link-modeling}
To determine communication times along these paths, TriMedium derives
link distances from node positions in the shared world. For acoustic
links, propagation loss combines geometric spreading and
frequency-dependent absorption~\cite{stojanovic2009acoustic}. The spreading
component follows a logarithmic dependence on distance, while absorption
is evaluated using the Thorp empirical relation~\cite{thorp1967attenuation}
at the configured carrier frequency and accumulated over the propagation
distance. Acoustic travel
time is calculated from node separation using a sound speed of 1500~m/s.

RF links use a log-distance path-loss model~\cite{sun2016pathloss}, with
a reference loss and path-loss exponent describing attenuation with node
separation. Configured line-of-sight (LOS) and
non-line-of-sight (NLOS) conditions select the corresponding exponent,
with additional attenuation applied to NLOS links. RF propagation delay
is calculated from the same node separation using the electromagnetic
propagation speed.

The propagation losses enter a link budget with the configured transmit
power and receiver noise to determine the received signal-to-noise ratio
(SNR). SNR and bandwidth yield a Shannon-based rate estimate, capped by the
maximum link rate. This rate and the actual encoded payload size determine
per-hop transmission time. These models provide a common rate and timing
baseline for examining how communication conditions and data processing
affect observation delivery to embodied tasks.

Future extensions could incorporate acoustic multipath and spatially
varying sound speeds to represent more complex underwater conditions. For
RF links, scene geometry could support obstacle-aware LOS/NLOS
determination, extending the current configuration-based propagation
conditions. Calibration against measured channels could further adapt the
propagation and rate models to specific environments and communication
hardware.

\subsection{Software Implementation}
\label{sec:software-implementation}
TriMedium runs on Linux with Unreal Engine 4.26 and a Python layer for
processing, communication, and receiver-side tasks. Configuration files
specify sampling, processing methods, and link parameters, allowing
experiments and processing algorithms to change without rebuilding the
Unreal runtime.

The ten-image reporting case (Section~\ref{sec:multimodal-delivery}) was run
with $640\times360$ RGB observations, using one NVIDIA GeForce RTX 4090 GPU
for rendering. The complete run covered 313.4~s of simulated observation
time and completed all four reference configurations in 447.6~s of
wall-clock time.

\section{Case Studies}
\label{sec:evaluation}

We examine TriMedium through three cross-medium embodied tasks. Telemetry
and image reporting tests complete end-to-end delivery, while
Air--Ground--Sea mapping examines whether observations from the three
domains can form a spatially consistent map. Shore-side AUV tracking then
examines how delayed measurements support position estimation for a moving
target.

\subsection{Multimodal Data Delivery}
\label{sec:multimodal-delivery}
We consider underwater observation reporting, in which a moving AUV sends
state telemetry and RGB images to a shore receiver through a surface
gateway and a UAV relay. Telemetry and images are evaluated in separate
experiments using the same acoustic--RF delivery path. Four reference
configurations exercise the processing interfaces: direct transmission,
lossless compression, and quantization followed by compression at two
precision levels. Within each experiment, these configurations receive
identical observations and motion histories. Table~\ref{tab:settings}
summarizes the settings.

\begin{table}[!t]
\caption{Telemetry delivery and reconstruction results.}
\label{tab:telemetry-results}
\centering
\footnotesize
\setlength{\tabcolsep}{2.5pt}
\renewcommand{\arraystretch}{1.12}
\begin{tabular*}{\columnwidth}{@{\extracolsep{\fill}}lrrrr@{}}
\toprule
\raisebox{4.3pt}{\textbf{Metric}}
  & \raisebox{4.3pt}{\textbf{FP32}}
  & \shortstack{\textbf{FP32}\\\textbf{+ zlib}}
  & \shortstack{\textbf{16-bit}\\\textbf{+ zlib}}
  & \shortstack{\textbf{8-bit}\\\textbf{+ zlib}} \\
\midrule
Volume (B) & 19,856 & 10,401 & 7,628 & 5,395 \\
\midrule
Processing (ms) & 0.443 & 1.328 & 2.655 & 2.655 \\
Transmission (ms) & 293.428 & 153.704 & 112.725 & 79.726 \\
Other (ms) & 70.727 & 70.727 & 70.727 & 70.727 \\
Total latency (ms) & 364.598 & 225.759 & 186.108 & 153.109 \\
\midrule
Max. position error (mm) & 0.000 & 0.000 & 0.298 & 82.024 \\
Max. velocity error (mm/s) & 0.000 & 0.000 & 0.048 & 11.901 \\
Max. depth error (mm) & 0.000 & 0.000 & 0.114 & 29.020 \\
\bottomrule
\end{tabular*}
\par\smallskip
\begin{minipage}{\columnwidth}
\footnotesize
Latencies are group means. Volume includes application-packet headers.
Latency components and totals are rounded independently.
\end{minipage}
\end{table}

\begin{table}[!t]
\caption{Image delivery and reconstruction results.}
\label{tab:image-results}
\centering
\footnotesize
\setlength{\tabcolsep}{2.5pt}
\renewcommand{\arraystretch}{1.12}
\begin{tabular*}{\columnwidth}{@{\extracolsep{\fill}}lrrrr@{}}
\toprule
\raisebox{4.3pt}{\textbf{Metric}}
  & \raisebox{4.3pt}{\textbf{PNG}}
  & \raisebox{4.3pt}{\textbf{zlib}}
  & \shortstack{\textbf{6-bit}\\\textbf{+ zlib}}
  & \shortstack{\textbf{4-bit}\\\textbf{+ zlib}} \\
\midrule
Volume (B) & 1,242,718 & 913,587 & 322,070 & 107,713 \\
\midrule
Latency at 0~s (s) & 30.43 & 22.27 & 8.38 & 3.25 \\
Latency at 36~s (s) & 100.02 & 68.13 & 8.33 & 3.01 \\
Latency at 72~s (s) & 152.73 & 95.61 & 8.39 & 3.00 \\
Latency at 108~s (s) & 205.38 & 122.82 & 8.33 & 2.98 \\
\midrule
RGB PSNR (dB) & / & / & 48.08 & 35.77 \\
Max. RGB error & 0 & 0 & 2 & 8 \\
\bottomrule
\end{tabular*}
\par\smallskip
\begin{minipage}{\columnwidth}
\footnotesize
Capture times are relative to the first selected frame; / denotes exact
RGB reconstruction. The zlib reference compresses decoded pixels.
\end{minipage}
\end{table}

\begin{figure}[!t]
\centering
\includegraphics[width=\columnwidth]{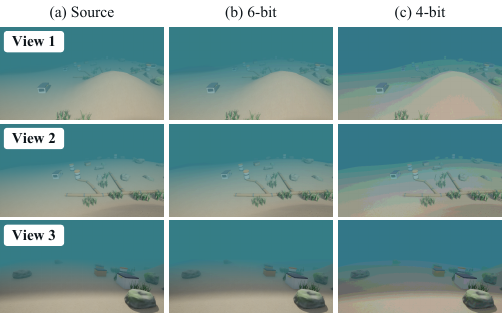}
\caption{Qualitative comparison of (a) source observations and shore-side
reconstructions using (b) 6-bit and (c) 4-bit uniform quantization with
zlib compression. Each row uses the same source frame.}
\label{fig:visual-outputs}
\end{figure}

In the telemetry and image experiments, all four configurations complete
end-to-end reporting, reconstructing 569 telemetry samples and ten images
at shore. Lossless configurations preserve the originals, while quantized
configurations reconstruct observations at the selected precision
(Fig.~\ref{fig:visual-outputs}).

\begingroup
\clubpenalty=10000
At a 12-s reporting interval, PNG transmission accumulates delay as new
images arrive before earlier transfers finish. Quantization plus
compression alleviates this backlog, while lower precision introduces
visible banding. Tables~\ref{tab:telemetry-results}
and~\ref{tab:image-results} report the corresponding data volumes,
latencies, and reconstruction errors.
\par\endgroup

\subsection{Air--Ground--Sea Mapping}
\label{sec:three-domain-mapping}
This case examines whether Air--Ground--Sea observations can form a
common coastal map. The AUV collects RGB images and rendered depth along
its trajectory, while aerial and ground cameras capture complementary
views at fixed survey poses. Known sensor poses place all RGB-D
observations in the shared world frame.

\begin{figure}[!t]
\centering
\includegraphics[width=\columnwidth]{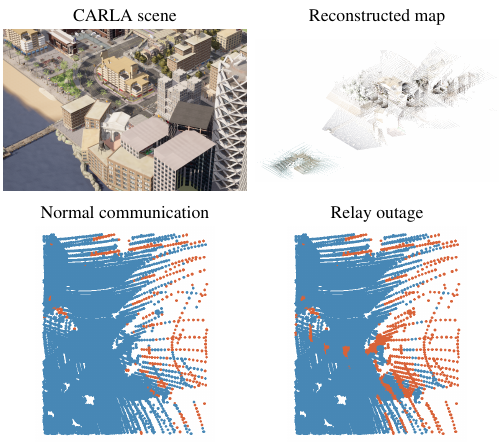}
\caption{Scene context and shore-side mapping results. Top: the scene view
and reconstructed RGB point-cloud map. Bottom: the same local underwater
reference region under normal communication and a relay outage at
40--55~s. Blue points have a map neighbour within 1~m; orange points do not.
The scene view is horizontally mirrored for display.}
\label{fig:mapping-case-study}
\end{figure}

Local point clouds projected from RGB-D data are compressed and
transmitted to shore, where only delivered observations are fused into a
0.4-m voxel map. Using identical source observations, we compare normal
communication with a UAV-relay outage at 40--55~s.

\begin{table}[!t]
\caption{Mapping quality against held-out reference observations.}
\label{tab:mapping-quality}
\centering
\footnotesize
\setlength{\tabcolsep}{3pt}
\renewcommand{\arraystretch}{1.12}
\begin{tabular*}{\columnwidth}{@{\extracolsep{\fill}}llrr@{}}
\toprule
\textbf{Region} & \textbf{Condition} & \textbf{Coverage (\%)} & \textbf{RMSE (m)} \\
\midrule
\multirow{2}{*}{All domains} & Normal & 87.32 & 0.149 \\
 & Relay outage & 86.91 & 0.142 \\
\midrule
\multirow{2}{*}{Underwater} & Normal & 93.08 & 0.313 \\
 & Relay outage & 91.24 & 0.308 \\
\bottomrule
\end{tabular*}
\par\smallskip
\begin{minipage}{\columnwidth}
\footnotesize
Coverage is the fraction of reference points within 1~m of the map. RMSE
uses map-to-reference nearest-point distances, capped at 1~m. Each region
is evaluated using all its held-out reference points.
\end{minipage}
\end{table}

All 32 observation groups arrive under normal communication. Coverage and
distance errors are evaluated against 32 independently captured, held-out RGB-D views
(Table~\ref{tab:mapping-quality}): coverage within 1~m is 87.32\% overall
and 93.08\% underwater. During the outage, two groups fail to arrive, leaving 30 for fusion, and
underwater coverage drops to 91.24\%
(Fig.~\ref{fig:mapping-case-study}). The resulting map reflects both
cross-domain observations and their successful delivery.

\subsection{Cross-Domain AUV Positioning}
\label{sec:auv-positioning}
This case study considers shore-side tracking of a moving AUV using
acoustic measurements obtained at a surface reference node. A geometric
measurement model generates range, azimuth, and elevation observations at
1~Hz, with range and angular noise standard deviations of 0.15~m and
$0.5^\circ$, respectively. Acoustic round-trip propagation and instrument
processing determine when measurements become available at the reference
node, after which they are reported through a UAV relay to shore over RF
links.

\begin{figure}[!t]
\centering
\includegraphics[width=\columnwidth]{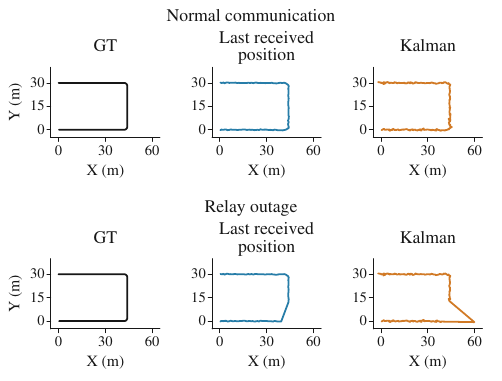}
\caption{Shore-side AUV positioning under normal communication (top) and
a relay outage at 40--55~s (bottom). Columns show ground truth (GT),
positions held until the next report, and Kalman estimates for one fixed
measurement-noise realization. All panels use the same relative XY coordinates and spatial
scale. Lines join successive estimates.}
\label{fig:auv-positioning-case-study}
\end{figure}

Using the known reference-node pose, the receiver converts delivered range
and angular measurements into AUV position observations. We compare two
strategies. The first treats the last received position observation as the
current estimate, holding it unchanged until a new report arrives. The
second uses a constant-velocity Kalman filter: on report reception, it
updates the motion estimate for the measurement acquisition time, then
predicts to the current simulation time. Both use the same delivered
measurements.

\begin{table}[!t]
\caption{Three-dimensional positioning RMSE (m).}
\label{tab:positioning-results}
\centering
\footnotesize
\setlength{\tabcolsep}{4pt}
\renewcommand{\arraystretch}{1.12}
\begin{tabular*}{\columnwidth}{@{\extracolsep{\fill}}lrr@{}}
\toprule
\textbf{Reporting condition} & \textbf{Last received position} & \textbf{Kalman} \\
\midrule
Normal communication & $0.807\pm0.014$ & $0.523\pm0.020$ \\
Relay outage & $2.896\pm0.022$ & $3.254\pm0.724$ \\
\bottomrule
\end{tabular*}
\par\smallskip
\begin{minipage}{\columnwidth}
\footnotesize
Mean $\pm$ sample standard deviation of per-realization RMSE across five
measurement-noise realizations on the same AUV trajectory.
\end{minipage}
\end{table}

Across five independent measurement-noise realizations on the same trajectory, mean
3D position RMSEs under normal communication are 0.807~m for retaining the last
position and 0.523~m for Kalman prediction
(Table~\ref{tab:positioning-results}). An outage at 40--55~s coincides
with an AUV turn: the retained position stays fixed, while the predictor
continues along the previous direction
(Fig.~\ref{fig:auv-positioning-case-study}), increasing RMSEs to 2.896~m
and 3.254~m, respectively. This case demonstrates how TriMedium exposes
the joint effects of communication availability and vehicle motion on
tracking.

Together, the three case studies demonstrate TriMedium's support for
observation reporting, cross-domain mapping, and AUV tracking, while
showing how communication conditions affect the information available to
receiving tasks. Building on these capabilities, future work will extend
the framework toward joint communication and computation design.
Observation processing could be distributed among AUVs, gateways, aerial
relays, and shore nodes, enabling studies of task offloading and resource
allocation~\cite{mao2017mec} along the cross-medium path. Processing placement and
transmission scheduling could then be coordinated according to link
conditions, computing capacity, and the information needs of mapping and
positioning tasks.

\FloatBarrier
\balance
\section{Conclusion}
\label{sec:conclusion}

This paper presented TriMedium for studying cross-medium cooperation among
aerial, ground, and underwater platforms. Agents acquire observations in
the shared environment, and the resulting data pass through selectable
processing methods and acoustic--RF delivery paths to the receiver.
Receiving tasks use the delivered data for reconstruction and estimation,
connecting simulated sensing and communication with task execution. Three
case studies demonstrate TriMedium's support for end-to-end telemetry and
image reporting, Air--Ground--Sea mapping, and shore-side AUV tracking.
These capabilities provide an experimental basis for studying
communication and computation decisions in cooperative tasks.


\bibliographystyle{IEEEtran}
\bibliography{references}

\begin{thebibliography}{10}
\providecommand{\url}[1]{#1}
\csname url@samestyle\endcsname
\providecommand{\newblock}{\relax}
\providecommand{\bibinfo}[2]{#2}
\providecommand{\BIBentrySTDinterwordspacing}{\spaceskip=0pt\relax}
\providecommand{\BIBentryALTinterwordstretchfactor}{4}
\providecommand{\BIBentryALTinterwordspacing}{\spaceskip=\fontdimen2\font plus
\BIBentryALTinterwordstretchfactor\fontdimen3\font minus
  \fontdimen4\font\relax}
\providecommand{\BIBforeignlanguage}[2]{{%
\expandafter\ifx\csname l@#1\endcsname\relax
\typeout{** WARNING: IEEEtran.bst: No hyphenation pattern has been}%
\typeout{** loaded for the language `#1'. Using the pattern for}%
\typeout{** the default language instead.}%
\else
\language=\csname l@#1\endcsname
\fi
#2}}
\providecommand{\BIBdecl}{\relax}
\BIBdecl

\bibitem{zeng2016uav}
Y.~Zeng, R.~Zhang, and T.~J. Lim, ``Wireless communications with unmanned
  aerial vehicles: Opportunities and challenges,'' \emph{IEEE Commun. Mag.},
  vol.~54, no.~5, pp. 36--42, May 2016.

\bibitem{liang2026synergetic}
H.~Liang, Y.~Diao, Y.~Wu, F.~Zhou, and Q.~Wu, ``Synergetic empowerment:
  Wireless communications meets embodied intelligence,'' \emph{IEEE Commun.
  Mag.}, pp. 1--8, 2026, early access.

\bibitem{carlaair2026}
T.~Zeng, Y.~Wen, and H.~Zhang, ``{CARLA-Air}: Fly drones inside a {CARLA} world
  -- a unified infrastructure for air-ground embodied intelligence,''
  arXiv:2603.28032, 2026.

\bibitem{gao2026lowaltitudeiscc}
S.~Gao \emph{et~al.}, ``Integrated sensing, communication, and computation for
  low-altitude networks towards seamless connectivity and connected
  intelligence,'' \emph{IEEE Internet Things Mag.}, vol.~9, no.~3, pp. 63--71,
  May 2026.

\bibitem{zhang2022dave}
M.~M. Zhang \emph{et~al.}, ``{DAVE} aquatic virtual environment: Toward a
  general underwater robotics simulator,'' in \emph{Proc. IEEE/OES Auton.
  Underwater Vehicles Symp. (AUV)}, Sep. 2022, pp. 1--8.

\bibitem{potokar2022holoocean}
E.~Potokar, S.~Ashford, M.~Kaess, and J.~G. Mangelson, ``{HoloOcean}: An
  underwater robotics simulator,'' in \emph{Proc. IEEE Int. Conf. Robot. Autom.
  (ICRA)}, May 2022, pp. 3040--3046.

\bibitem{cieslak2019stonefish}
P.~Cie{\'s}lak, ``{Stonefish}: An advanced open-source simulation tool designed
  for marine robotics, with a {ROS} interface,'' in \emph{Proc.
  OCEANS--Marseille}, Jun. 2019, pp. 1--6.

\bibitem{yan2026simart}
K.~Yan, Y.~Cao, J.~Li, L.~Xiang, and K.~Yang, ``{SimART}: A unified and open
  real-world multimodal simulation platform for {6G} integrated sensing and
  communication,'' arXiv:2605.13309, 2026.

\bibitem{hoydis2023sionnart}
J.~Hoydis \emph{et~al.}, ``{Sionna RT}: Differentiable ray tracing for radio
  propagation modeling,'' in \emph{Proc. IEEE GLOBECOM Workshops (GC Wkshps)},
  Dec. 2023, pp. 317--321.

\bibitem{riley2010ns3}
G.~F. Riley and T.~R. Henderson, ``The {ns-3} network simulator,'' in
  \emph{Modeling and Tools for Network Simulation}.\hskip 1em plus 0.5em minus
  0.4em\relax Berlin, Germany: Springer, 2010, pp. 15--34.

\bibitem{dosovitskiy2017carla}
A.~Dosovitskiy, G.~Ros, F.~Codevilla, A.~Lopez, and V.~Koltun, ``{CARLA}: An
  open urban driving simulator,'' in \emph{Proc. Conf. Robot Learn. (CoRL)},
  vol.~78, Nov. 2017, pp. 1--16.

\bibitem{shah2018airsim}
S.~Shah, D.~Dey, C.~Lovett, and A.~Kapoor, ``{AirSim}: High-fidelity visual and
  physical simulation for autonomous vehicles,'' in \emph{Field and Service
  Robotics}, ser. Springer Proceedings in Advanced Robotics, vol.~5.\hskip 1em
  plus 0.5em minus 0.4em\relax Springer, 2018, pp. 621--635.

\bibitem{amer2023unav}
A.~Amer, O.~{\'A}lvarez-Tu{\~n}{\'o}n, H.~{\.I}. U{\u{g}}urlu, J.~{Le Fevre
  Sejersen}, Y.~Brodskiy, and E.~Kayacan, ``{UNav-Sim}: A visually realistic
  underwater robotics simulator and synthetic data-generation framework,'' in
  \emph{Proc. 21st Int. Conf. Adv. Robot. (ICAR)}, Dec. 2023, pp. 570--576.

\bibitem{alqurashi2023maritime}
F.~S. Alqurashi, A.~Trichili, N.~Saeed, B.~S. Ooi, and M.-S. Alouini,
  ``Maritime communications: A survey on enabling technologies, opportunities,
  and challenges,'' \emph{IEEE Internet Things J.}, vol.~10, no.~4, pp.
  3525--3547, Feb. 2023.

\bibitem{wubben2023cosimulation}
J.~Wubben, C.~T. Calafate, F.~Granelli, J.-C. Cano, and P.~Manzoni, ``A
  real-time co-simulation framework for multi-{UAV} environments offering
  detailed wireless channel models,'' in \emph{Proc. IEEE Int. Conf. Commun.
  (ICC)}, May 2023, pp. 2649--2654.

\bibitem{stojanovic2009acoustic}
M.~Stojanovic and J.~Preisig, ``Underwater acoustic communication channels:
  Propagation models and statistical characterization,'' \emph{IEEE Commun.
  Mag.}, vol.~47, no.~1, pp. 84--89, Jan. 2009.

\bibitem{thorp1967attenuation}
W.~H. Thorp, ``Analytic description of the low-frequency attenuation
  coefficient,'' \emph{J. Acoust. Soc. Am.}, vol.~42, no.~1, p. 270, Jul. 1967.

\bibitem{sun2016pathloss}
S.~Sun \emph{et~al.}, ``Propagation path loss models for {5G} urban micro- and
  macro-cellular scenarios,'' in \emph{Proc. IEEE 83rd Veh. Technol. Conf. (VTC
  Spring)}, May 2016, pp. 1--6.

\bibitem{mao2017mec}
Y.~Mao, C.~You, J.~Zhang, K.~Huang, and K.~B. Letaief, ``A survey on mobile
  edge computing: The communication perspective,'' \emph{IEEE Commun. Surveys
  Tuts.}, vol.~19, no.~4, pp. 2322--2358, 2017.

\end{thebibliography}

\end{document}